\documentclass[prb,aps,twocolumn,showpacs,preprintnumbers,amsmath,amssymb]{revtex4-2}
\usepackage{amsmath}
\usepackage{graphicx}
\usepackage{dcolumn}
\usepackage{bm}
\usepackage{float}
\usepackage{xcolor}
\usepackage{graphicx}
\usepackage{booktabs}
\usepackage{multirow}
\usepackage[colorlinks=true,linkcolor=blue]{hyperref}
\begin{document}
\title{Dopant Cluster Engineering for Tunable Room-Temperature Ferromagnetism in 2D Transition Metal Dichalcogenides}
\author{Shivangee}
\author{Sachin Grewal}
\author{Akash Singh}
\email {akashsingh@iitp.ac.in}
\affiliation{Department of Physics, Indian Institute of Technology, Patna 801106, India}
\date{\today}
\begin{abstract}
Two-dimensional transition metal dichalcogenides (2D-TMDs) offer a promising platform for realizing dilute magnetic semiconductors. However, simultaneously achieving large magnetic anisotropy energy (MAE) and high Curie temperature (T$_\text{c}$) remains a major challenge. In this work, we employ DFT+U to systematically investigate the thermodynamic stability and magnetic properties of 2D-MoSe$_2$ and 2D-WSe$_2$ doped with transition metal atoms. We found that the paired dopant configurations are more stable than previously attributed isolated dopants. Despite their nearly identical crystal structures, 2D-MoSe$_2$ and 2D-WSe$_2$ show contrasting properties upon doping. All paired dopants in 2D-MoSe$_2$ exhibit long-range ferromagnetic coupling (FMC) primarily due to carrier-mediated exchange interactions, whereas those in 2D-WSe$_2$ show weak exchange coupling. Interestingly, 2Mn$_\text{2W}$ shows a  transition from less stable antiferromagnetic coupling (AFMC) in the short range to highly stable FMC in the long-range. The combined effect of spin-orbit coupling, exchange interactions, and orbital hybridization gives rise to a large MAE with maximum in-plane values of $-$3.5 meV for 2V$_\text{2Mo}$ and $-$3.2 meV for 2Mn$_\text{2W}$.  Monte Carlo simulations estimate T$_\text{c}$ $\simeq$ 440 K for 2Fe$_\text{2Mo}$ and ensure its high-temperature stability in 2D-MoSe$_2$. Our results establish dopant clustering as an effective route to engineer high-temperature ferromagnetism in 2D-TMDs and highlight their potential for nano-spintronic applications.\end{abstract}
\maketitle

\section{INTRODUCTION}
\label{Introduction}

Two-dimensional (2D) magnetic materials are attracting strong interest in quantum spintronics as they provide a platform for exploring magnetism at the atomic scale. Their atomically thin nature offers high information storage density, tunable magnetic properties, and seamless integration with other 2D materials \cite{ahn20202d,elahi2022review,zhang20242d,singh2018engineering}.  However, according to the Mermin-Wagner theorem \cite{mermin1966absence}, realizing magnetism in low-dimensional systems is inherently challenging due to thermal fluctuations. Fortunately, magnetic anisotropy, which originates mainly from spin-orbit coupling (SOC) \cite{wang1993first,lado2017origin}, can overcome this restriction and promote long-range magnetism in practical 2D materials. A major breakthrough came in 2017 with the successful synthesis of intrinsic 2D ferromagnetic semiconductors such as CrI$_\text{3}$ \cite{huang2017layer} and Cr$_\text{2}$Ge$_\text{2}$Te$_\text{6}$ \cite{gong2017discovery}. These discoveries proved that magnetism can exist in atomically thin materials. So far, several intrinsic magnetic semiconductors have been found to exhibit sizable magnetic anisotropy; however, their Curie temperatures remain far below room temperature and limit their practical applicability \cite{huang2017layer,gong2017discovery,cai2019atomically,zhang2019direct,lee2021magnetic,achinuq2022covalent}. Therefore, finding a magnetic 2D material with large magnetic anisotropy and high Curie temperature remains a key challenge in materials design. 
 
  One way to address this challenge is by doping conventional semiconductors with magnetic atoms to create dilute magnetic semiconductors (DMSs) \cite{munekata1989diluted,shen1997ga,sato2010first,singh2022magnetic}. This approach allows materials to retain their semiconducting properties while gaining magnetic functionality \cite{mishra2013long,ramasubramaniam2013mn,habib2018ferromagnetism,yang2019ta}. In particular, in-situ substitutional doping is known to yield structurally stable DMSs \cite{loh2021substitutional}. Among potential host materials, transition metal dichalcogenides (TMDs) are especially promising due to the ease of incorporating transition metal (TM) impurities \cite{shu2015layer}, and the tunability of electronic and magnetic properties by dopant selection \cite{liu2024modulation,singh2021atypical}. TM impurities such as V, Mn, Fe, and Co are readily incorporated into these materials during growth \cite{fang2023advances}, making them highly viable candidates for DMSs. Moreover, the partially occupied $d$ orbitals of TM atoms in TMDs facilitate the induction of magnetism through doping \cite{PhysRevB.88.245436}. Additionally, their direct band gaps in the visible-near-infrared range \cite{Zeng2013,zhang2014direct} make them highly desirable for spintronics as they enable optical generation and control of spin-polarized carriers \cite{wang2018colloquium,singh2019origin}. Despite these advantages, a systematic understanding of magnetic ordering, magnetic anisotropy, and thermodynamic stability in $4d$ (or $5d$) TMD based DMSs is still lacking, and their suitability as host materials for DMS remains largely unexplored. 
  
\begin{figure*}
\centering
\includegraphics[width=2\columnwidth]{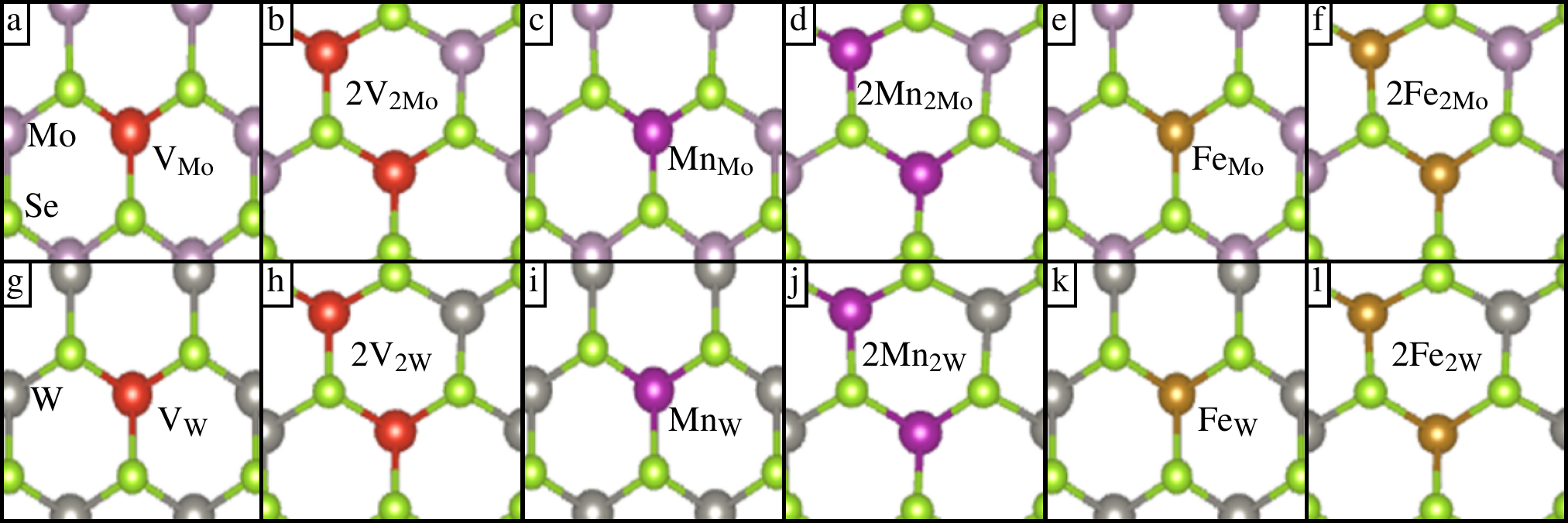}
\caption{Top view of the fully relaxed structures of isolated and paired transition metal (TM) dopants in 2D-MoSe$_\text{2}$ and 2D-WSe$_2$. Panels (a-f) correspond to V, Mn, and Fe atoms substituting Mo sites in MoSe$_2$, while panels (g-l) show the corresponding substitutions at W sites in WSe$_2$. TM$_\text{Mo (or W)}$ (TM$=$V, Mn, or Fe) represents a single TM atom substituting a Mo (or W) atom, whereas 2TM$_\text{2Mo (or 2W)}$ represents two TM atoms substituting two Mo (or W) sites. Here, the Se atoms are shown in neon green, Mo in light purple, W in grey, V in dark orange, Mn in dark magenta, and Fe in golden brown.}
\label{fig:comparison-str}
\end{figure*}
  
  Here, using DFT$+$U, we have performed a comprehensive study on the thermodynamic, electronic, and magnetic properties of $3d$ TM (i.e., V, Mn, and Fe) doped 2D-MoSe$_2$ and 2D-WSe$_2$. The defect thermodynamic analysis suggests that these TM elements prefer pairing or clustering, i.e., TM$_\text{Mo}$-TM$_\text{Mo}$ (or TM$_\text{W}$-TM$_\text{W}$) rather than being well separated, isolated TM$_\text{Mo}$ (or TM$_\text{W}$) defects under Se-rich growth conditions. Most of these dopant clusters provide robust long-range ferromagnetism, which arises mainly due to carrier-mediated exchange interactions of Ruderman-Kittel-Kasuya-Yoshida (RKKY) \cite{sato2010first} type in which the localized magnetic moments at the TM sites polarize the itinerant charge carriers. Moreover, there is a superexchange interaction between the two TM atoms mediated via Se atoms. To ensure the thermal stability of the spin-polarized states, we have calculated the magnetic anisotropy energy (MAE) along with its microscopic origin for all the doped systems. These doped systems exhibit large MAEs, with 2V$_\text{2Mo}$ and 2Mn$_\text{2W}$ showing in-plane MAE of $-$3.5 meV and $-$3.2 meV, respectively. Monte Carlo simulations estimate Curie temperatures (T$_\text{c}$) in the range of $\simeq$ 340K to 440 K for the TM$_\text{Mo}$-TM$_\text{Mo}$ (or 2TM$_\text{2Mo}$). These results establish dopant clustering as an effective route to realize 2D-TMDs with robust magnetism, large magnetic anisotropy and high T$_\text{c}$, making them promising for next-generation quantum spintronic applications.
\section{METHODOLOGY}
\label{develop}
All calculations are performed using density functional theory (DFT) with the projector augmented wave (PAW) method \cite{blochl1994projector,kresse1999ultrasoft}, as implemented in Vienna Ab initio Simulation Package (VASP) \cite{kresse1993ab,kresse1994ab}. The exchange correlation potential is treated using the generalized gradient approximation (GGA) with Perdew-Burke-Ernzerhof (PBE) functional \cite{perdew1996generalized}. Spin-polarized DFT$+$U calculations are performed to account for the on-site Coulomb interactions between $d$ electrons of the TM atoms \cite{dudarev1998electron,himmetoglu2014hubbard}. Hubbard U values of 2, 3 and 4 eV are examined and U $=$ 3 eV is found to produce results close to experimental values. Consequently, U $=$ 3 eV is used in all the calculations. A supercell approach is adopted to study the effects of doping TM atoms in 2D-MoSe$_\text{2}$ and 2D-WSe$_\text{2}$. To eliminate spurious electrostatic interactions between the dopants, a sufficiently large 5$\times$5$\times$1 supercell is used. A $\Gamma$-centered 2$\times$2$\times$1 $k$ mesh is employed for the supercell. The plane-wave cutoff energy is set to 400 eV, with a total energy convergence threshold of $10^{-4}$ eV. Structural relaxations are considered complete when the Hellmann-Feynman force on each atom falls below 0.02 eV/\AA. A vacuum layer greater than 15~\AA\ is introduced along the z-axis to avoid artificial interactions between periodic images. A single TM (V, Mn, or Fe) atom is substituted at a Mo (or W) site in the 5$\times$5$\times$1 supercell, corresponding to a doping concentration of 1.3\%, representing an isolated dopant i.e., TM$_\text{Mo (or W)}$. To investigate the experimentally observed dopant clustering, two Mo (or W) atoms are replaced by TM atoms i.e., 2TM$_\text{2Mo (or 2W)}$, which results in a doping concentration of 2.6\%.

\begin{figure*}[!ht]
\centering
\includegraphics[width=2\columnwidth]{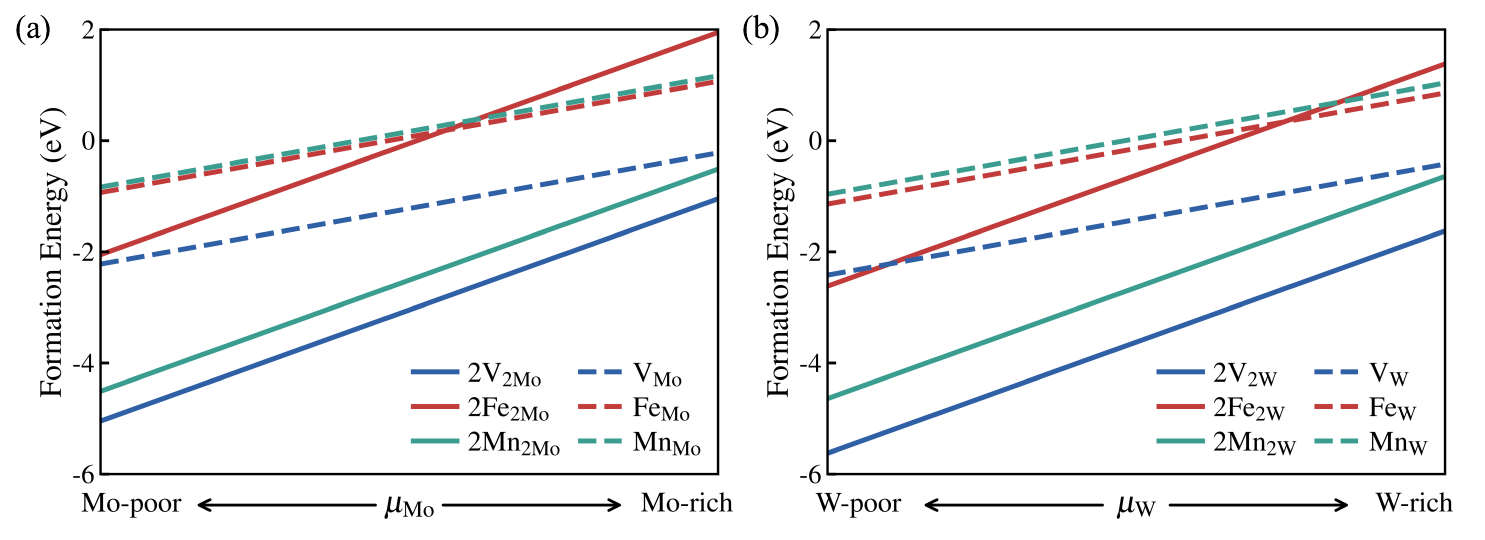}
\caption{Formation Energy (E$^{f}$) of V, Mn, and Fe substitution in (a) 2D-MoSe$_\text{2}$ and (b) 2D-WSe$_\text{2}$ as a function of Mo (or W) chemical potential $\mu_\mathrm{Mo (or W)}$, under growth conditions ranging from Mo (or W)-poor to Mo (or W)-rich limits. Dashed lines correspond to substitution of a single dopant (TM$_\text{Mo (or W)}$), while the solid lines correspond to paired dopants (2TM$_\text{2Mo (or 2W)}$).}
\label{fig:E_fvsmu_Mo(W)}
\end{figure*}

The thermodynamic stability of doped systems has been analyzed using the formation energy under Mo (or W)-rich and Mo (or W)-poor conditions for both TM$_\text{Mo (or W)}$ and 2TM$_\text{2Mo (or 2W)}$. The formation energy ($\text{E}^{f}$) of a dopant is defined as
$$
{\text{E}^{f} = \mathrm{E_{doped} - E_{pristine} -}} n\mathrm{(\mu_{TM } - \mu_{Mo (or W)})}.
$$
Here, $\mathrm{E_{doped}}$ and $\mathrm{E_{pristine}}$ represent the total energies of fully relaxed doped and pristine supercells, respectively; $n$ is the number of substituted TM atoms, while $\mu_{\text{TM}}$ and $\mu_{\text{Mo (or W)}}$ are the chemical potentials of the TM dopant and host, Mo (or W) atoms. 
The chemical potentials of V, Mn, and Fe are calculated from the total energy per atom of their most stable bulk phases. For Mo and W, the chemical potentials are evaluated under two limiting growth conditions. In the Mo (or W)-rich limit, $\mu_{\text{Mo (or W)}}$ is derived from the energy per atom of the respective bulk bcc structure. In the Mo (or W)-poor limit, the chemical potentials are determined using the thermodynamic equilibrium condition
$\mathrm{\mu_{Mo (or W)}=E_{MoSe_2 (or WSe_2)}-2\mu_{Se}}$ where,
$\mathrm{E_{MoSe_2 (or WSe_2)}}$ is the total energy per unit cell of 2D-MoSe$_2$ (or 2D-WSe$_2$). $\mu_{\text{Se}}$  is the chemical potential of selenium, obtained from its bulk hexagonal phase.
The MAE is evaluated via the magnetic force theorem \cite{WANG1996337} by performing non-collinear calculations, where SOC interaction is introduced non-self-consistently as a perturbation to the scalar relativistic Hamiltonian. To understand the microscopic origin of the MAE, we analyzed the  SOC interactions between TM-$d$ states using second-order perturbation theory \cite{wang1993first}. Based on the calculated band energies and orbital-projected wavefunctions, we estimated the SOC matrix elements associated with the dopant $d$-orbitals.
Next, statistical methods are employed to determine the magnetic phase transition temperature. Monte Carlo simulations based on the Ising model \cite{onsager1944crystal} have been used to estimate Curie and N\'eel temperatures. A large 50$\times$50$\times$1 supercell with periodic boundary conditions is used to minimize finite-size effects. The Metropolis algorithm \cite{metropolis1953equation} is utilized with $10^5$ steps to achieve thermal equilibrium, which is followed by $10^6$ sampling steps at each temperature. Thermodynamic observables, including magnetization, specific heat, and magnetic susceptibility, are averaged over these configurations and normalized per spin. 
\enlargethispage{\baselineskip}
\section{RESULTS AND DISCUSSION}
\label{results}

\begin{figure}[!b]
\centering
\includegraphics[width=\columnwidth]{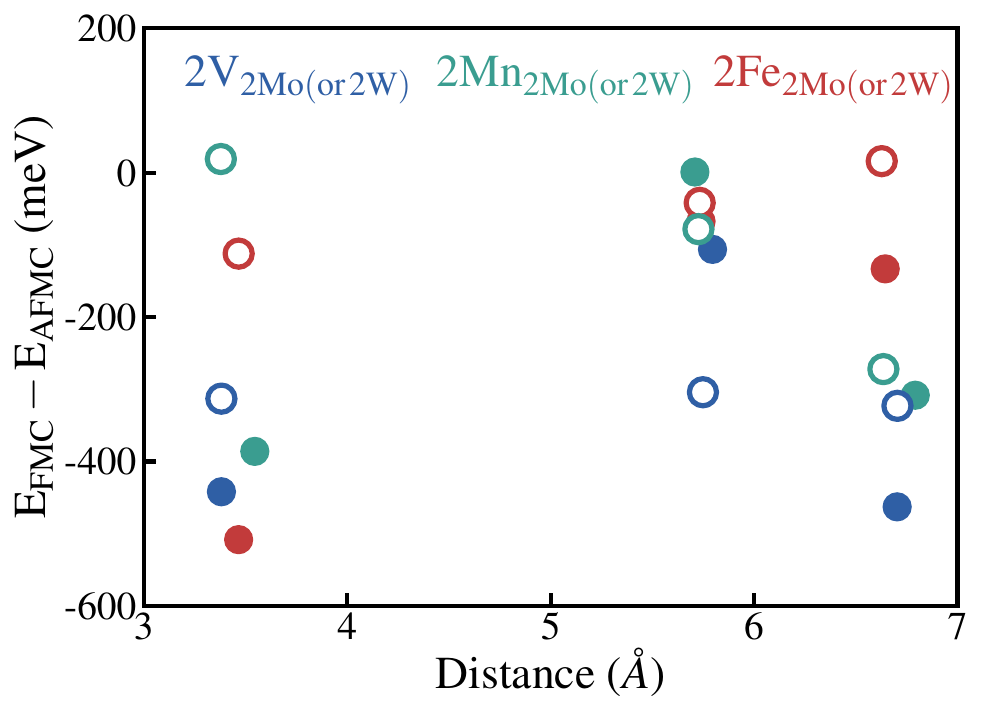}
\caption{Energy difference between the FMC and AFMC configurations ($\mathrm{\Delta E = E_{FMC} - E_{AFMC}}$) as a function of dopant-dopant separation, where the filled circles represent $\Delta \text{E}$ for 2D-MoSe$_\text{2}$, while the open circles represent that for 2D-WSe$_\text{2}$. Negative and positive values of $\Delta \text{E}$ indicate stability of the FMC and AFMC configurations, respectively.}
 \label{fig:fm-afm-vs-distance}
\end{figure}

\begin{figure*}[!ht]
\centering
\includegraphics[width=2\columnwidth]{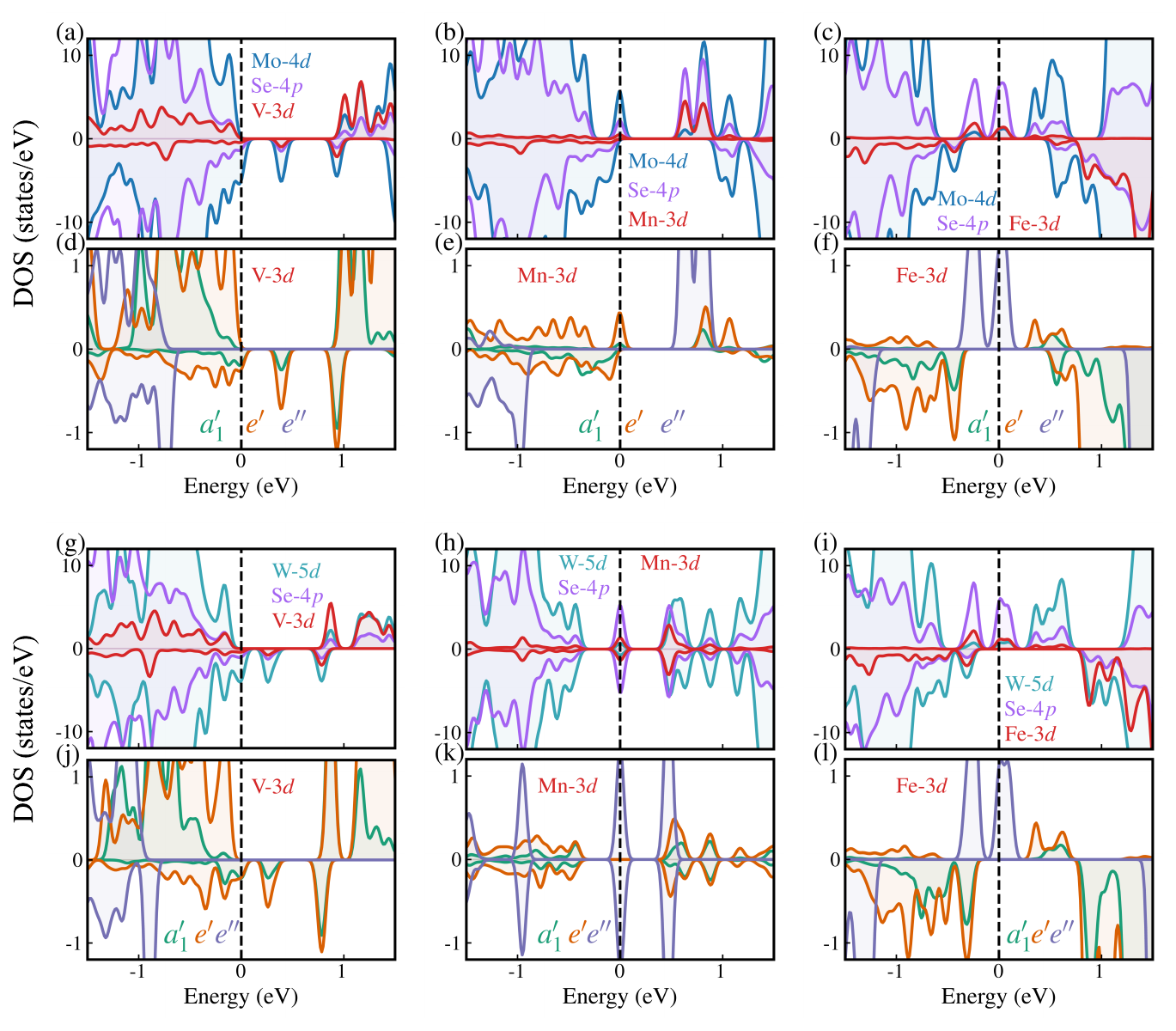}
\caption{Spin-polarized projected density of states (PDOS) of TM (V, Mn, Fe) doped 2D-MoSe$_2$ and 2D-WSe$_2$. Positive and negative DOS values correspond to spin-up and spin-down states, respectively. Panels (a$-$c) show the Mo-$4d$ (sky blue), Se-$4p$ (violet), and dopant TM-$3d$ (dark red) orbital contributions for $\mathrm{2V_{2Mo}}$, $\mathrm{2Mn_{2Mo}}$, and $\mathrm{2Fe_{2Mo}}$ in 2D-MoSe$_\text{2}$, while panels (d$-$f) show the corresponding TM-$3d$ orbital resolved PDOS. Panels (g$-$i) and (j$-$l) present the analogous results for 2D-WSe$_2$, showing the W-5$d$ (cyan), Se-$4p$, and TM-$3d$ orbital contributions and the corresponding TM-$3d$ orbital resolved PDOS for $\mathrm{2V_{2W}}$, $\mathrm{2Mn_{2W}}$, and $\mathrm{2Fe_{2W}}$. The $a^{\prime}_{1}$, $e{'}$, and $e{''}$ states are shown in green, orange, and faded-violet, respectively. The energies are referenced with respect to the Fermi level (E$_\text{F}=0$), represented by the vertical dashed lines.}
\label{fig:dos}
\end{figure*}

 2D-MoSe$_\text{2}$ and 2D-WSe$_\text{2}$ crystallize in the hexagonal structure with space group P$\mathrm{6_3}/mmc$, where transition metal atoms (Mo or W) are sandwiched between two layers of chalcogen (Se) atoms~\cite{Chhowalla2013,doi:10.1021/nn400280c}. The optimized lattice parameters are $a=b=3.32$ ~\AA\ for both TMDs, with identical Mo-Se and W-Se bond lengths of 2.54~\AA, consistent with previous reports~\cite{huang2015alloy}. Their DFT$+$U calculated band gaps are direct at the $K$-point, with values of 1.42 eV (2D-MoSe$_\text{2}$) and 1.54 eV (2D-WSe$_\text{2}$), which agree well with experimental reports~\cite{tongay2014two,zhang2014two}. The spin-polarized partial density of states (PDOS) exhibits symmetric spin-up and spin-down states, confirming the non-magnetic behavior of these pristine TMDs. TM atoms in TMDs possess partially filled $d$ orbitals, enabling the possibility of inducing magnetism through doping. Here, we have doped 1.3\% concentration of isolated V, Mn, and Fe at the Mo (or W) site denoted by TM$_\text{Mo (or W)}$ in a 5$\times$5$\times$1 supercell of both 2D-TMDs, as shown in Fig.~\ref{fig:comparison-str}.
 To represent the dopant clustering, and analyze the effect of dopant separation on magnetic ordering, we have replaced a pair of host atoms with the considered magnetic impurities, 2TM$_\text{2Mo (or 2W)}$, resulting in their 2.6\% concentration in the supercell as shown in Fig.~\ref{fig:comparison-str}. After complete structural relaxation, the TM-Se bond lengths are found to be slightly anisotropic, ranging from 2.5 \AA~to 2.6\AA~, indicating small local distortions around the impurity sites. The small variation in bond lengths suggests that the host lattice accommodates the dopants without significant structural deformation. To evaluate the energetic stability of the different dopant configurations, we calculated the formation energies of both isolated dopants and dopant clusters from their respective magnetic ground state; the results are summarized in Fig.~\ref{fig:E_fvsmu_Mo(W)}. Lower E$^f$ indicates easier incorporation of the dopant under the corresponding growth conditions. Under Mo (or W)-poor conditions, all doping configurations i.e., TM$_\text{Mo (or W)}$ and 2TM$_\text{2Mo (or 2W)}$ exhibit negative E$^{f}$, indicating their unintentional incorporation during the synthesis of 2D-MoSe$_2$ and 2D-WSe$_2$.
Among these three dopants, V substitution exhibits the lowest E$^f$ under both Mo-poor and W-poor conditions, suggesting strong compatibility with the host lattice.
Overall, the dopant clusters are found to be energetically more favorable with significantly lower E$^f$ compared to isolated dopants in 2D-MoSe$_2$ and 2D-WSe$_2$. This finding is consistent with the experimental studies showing that TM impurities tend to form clusters in 2D-TMDs~\cite{zhou2020synthesis,munson2025influence}. Therefore, our study focuses on the electronic and magnetic properties of these dopant clusters to realize robust spintronic functionalities in 2D-TMDs.
 
 Substitutional doping induces finite magnetic moments, whose magnitude depends on the number of charge carriers contributed by these dopants and their interaction with surrounding atoms. The calculated total and dopant site-projected magnetic moments are listed in Table~\ref {tab:magnetic-moments}. These values indicate strong localized moments for most of the dopant configurations. The dopant atoms form covalent bonds with Se atoms, leading to charge transfer and hybridization between the TM-$3d$ and Se-$4p$ orbitals. 
Electrons in the $d$ orbitals favor parallel spin alignment, resulting in partially filled $d$ orbitals, inducing localized magnetic moments. These magnetic moments interact and stabilize themselves in parallel or antiparallel coupling, depending on dopant hybridization, dopant-dopant separation, and host charge carrier concentration. We have considered ferromagnetic coupling (FMC) and antiferromagnetic coupling (AFMC) when the magnetic moments of the two dopant atoms are aligned parallel and antiparallel, respectively.
\setlength{\heavyrulewidth}{0.08em}
\setlength{\lightrulewidth}{0.05em}
\setlength{\cmidrulewidth}{0.03em}
\begin{table}[!b]
\centering
\setlength{\tabcolsep}{5pt}
\begin{tabular}{llcccccc}
\toprule
\multirow{2}{*}{Material} & \multirow{2}{*}{Config.} 
& \multicolumn{3}{c}{$\mu_\text{dopant}$ ($\mu_\text{B}$)} 
& \multicolumn{3}{c}{$\mu_{\text{total}}$ ($\mu_\text{B}$)} \\
\cmidrule(lr){3-5} \cmidrule(lr){6-8}
 &  & V & Mn & Fe & V & Mn & Fe \\
\midrule
\multirow{3}{*}{2D-MoSe$_2$} 
 & 1NN & 1.6 & 3.4 & 3.4 & 2.1 & 2.5 & 3.7 \\
 & 2NN & 0.2 & 3.6 & 3.5 & 1.9 & 2.4 & 3.5 \\
 & 3NN & 0.02 & 3.6 & 3.4 & 1.9 & 2.4 & 3.8 \\
\midrule
\multirow{3}{*}{2D-WSe$_2$} 
 & 1NN & 1.7 & 3.7 & 3.4 & 1.9 & 0.0 & 3.8 \\
 & 2NN & 1.7 & 3.7 & 3.4 & 1.9 & 2.0 & 3.6 \\
 & 3NN & 1.6 & 3.8 & 3.4 & 1.9 & 2.0 & 0.0 \\
\bottomrule
\end{tabular}
\caption{Dopant site-projected magnetic moment ($\mu_\text{dopant}$) and total magnetic moment ($\mu_{\text{total}}$) of 2V$_\text{2Mo}$, 2Mn$_\text{2Mo}$, and 2Fe$_\text{2Mo}$ in 2D-MoSe$_\text{2}$ \& 2V$_\text{2W}$, 2Mn$_\text{2W}$, and 2Fe$_\text{2W}$ in (b) 2D-WSe$_\text{2}$, with dopants at 1NN, 2NN, and 3NN separations. The site-projected values represent the magnitude of the local magnetic moment.}
\label{tab:magnetic-moments}
\end{table} 
To examine the range of magnetic ordering, we have considered three dopant pair configurations: (a) 1NN, where dopants are placed at nearest neighbor position of the zigzag chain; (b) 2NN, where dopants are in second nearest neighbor position of the armchair chain; (c) 3NN, where dopants are placed at third nearest neighbor position of the zigzag chain. The magnetic ground state is determined from the total energy difference between FMC and AFMC configurations, $\mathrm{\Delta E = E_{FMC} - E_{AFMC}}$. The negative and positive values of $\Delta \text{E}$ indicate FMC and AFMC ground states, respectively. The calculated values of $\Delta \text{E}$ as a function of dopant separation are shown in Fig.~\ref{fig:fm-afm-vs-distance}.  Our results reveal that FMC is energetically more favorable than AFMC in nearly all the investigated TM-doped TMDs. However, 2Mn$_\text{2W}$ at the 1NN separation and 2Fe$_\text{2W}$ at the 3NN separation stabilize in AFMC by 19 meV and 16 meV, respectively. Interestingly, in 2D-MoSe$_\text{2}$, FMC is strongest at the 1NN position and weakens significantly at 2NN before recovering again at 3NN. This oscillatory magnetic behavior indicates the presence of Ruderman-Kittel-Kasuya-Yoshida (RKKY) indirect exchange interactions mediated by itinerant charge carriers. However, no clear oscillatory trend is observed in these TM-doped 2D-WSe$_\text{2}$. For instance, in 2Mn$_\text{2W}$, the magnetic exchange evolves from AFMC at 1NN to a progressively more stable FMC at 2NN and 3NN separations. 
For 2Fe$_\text{2W}$, the stability of FMC decreases progressively with increasing dopant separation, whereas 2V$_\text{2W}$ remains stable in FMC up to the 3NN separation. These distinct trends suggest that the exchange interactions in 2D-WSe$_\text{2}$ evolve differently from those in 2D-MoSe$_\text{2}$. The stronger SOC and enhanced anisotropic charge localization in 2D-WSe$_\text{2}$ may modify the oscillation period of the exchange interaction. Therefore, an oscillatory magnetic behavior may emerge at larger dopant separations. The predominance of FMC highlights the ability of these materials to support spin-polarized electronic states, which is particularly promising for spintronic devices.

\begin{figure}
\centering
\includegraphics[width=\columnwidth]{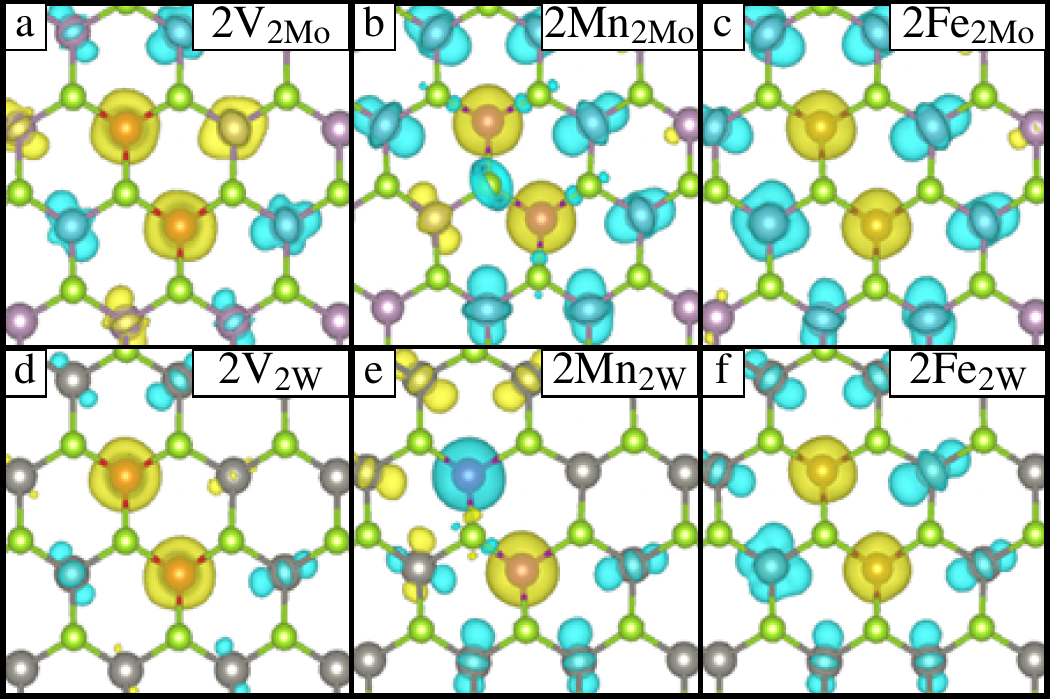}
\caption{Spin density distribution for $\mathrm{2V_{2Mo}}$, $\mathrm{2Mn_{2Mo}}$, and $\mathrm{2Fe_{2Mo}}$ in 2D-MoSe$_\text{2}$ (a$-$c), and for $\mathrm{2V_{2W}}$, $\mathrm{2Mn_{2W}}$, and $\mathrm{2Fe_{2W}}$ in 2D-WSe$_\text{2}$ (d$-$f). The isosurfaces are plotted at 0.01 e/\AA$^\mathrm{3}$. Yellow and cyan isosurfaces represent spin-up and spin-down charge densities, respectively, illustrating the magnetic coupling between the dopants, mediated by the host lattice.}
\label{fig:spin-density}
\end{figure}
\begin{figure*}[!ht]
\centering
\includegraphics[width=2\columnwidth]{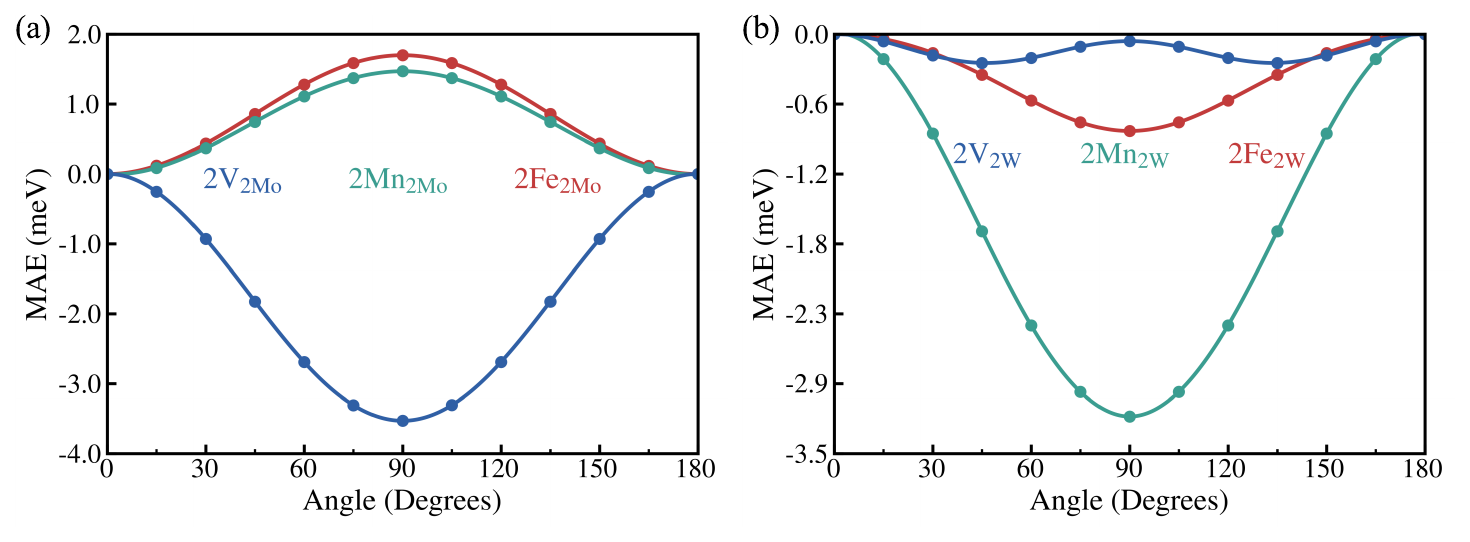}
\caption{Magnetic anisotropy energy (MAE) as a function of magnetization angle for $\mathrm{2V_{2Mo}}$, $\mathrm{2Mn_{2Mo}}$, and $\mathrm{2Fe_{2Mo}}$ in (a) 2D-MoSe$_2$ and for $\mathrm{2V_{2W}}$, $\mathrm{2Mn_{2W}}$, and $\mathrm{2Fe_{2W}}$ in (b) 2D-WSe$_2$. Positive values indicate perpendicular magnetic anisotropy (PMA), while negative values indicate an in-plane magnetic anisotropy (IMA).}
\label{fig:MAE-vs-angle}
\end{figure*}

 Next, to identify the origin of magnetic behavior, we systematically investigate the spin-polarized partial density of states (PDOS) of the doped systems up to 3NN separation in their respective magnetic ground states. The element resolved and TM-$d$ orbital resolved PDOS of 2TM$_\text{2Mo}$ and 2TM$_\text{2W}$ at 1NN separation are shown in Fig.~\ref{fig:dos}. The electronic states near the Fermi-level (E$_\text{F}$) exhibit strong hybridization of TM-$3d$, Mo-$4d$ (or W-$5d$), and Se-$4p$ orbitals, as shown in Fig. ~\ref{fig:dos}. In both 2D-MoSe$_{2}$ and 2D-WSe$_{2}$, the trigonal prismatic coordination of the surrounding chalcogen atoms gives rise to a $D_\text{3h}$ crystal field. Due to this, the five $d$-orbitals of the TM-atoms split into three groups: a non degenerate a$^{\prime}_{1}$ state ($d_\mathrm{z^2}$), a doubly degenerate e${'}$ state ($d_\mathrm{x^2-y^2}$, $d_\mathrm{xy}$), and a doubly degenerate e${''}$ state ($d_\mathrm{xz}$, $d_\mathrm{yz}$), as shown in the TM-$d$ orbital resolved PDOS [Fig.~\ref{fig:dos} (d-f), (j-l)]. In addition, the dopant $d$ states exhibit a pronounced exchange splitting, indicated by the energetic asymmetry between the spin-up and spin-down states. 2V$_\text{2Mo}$ induces hybridized spin-down states at E$_\text{F}$ with dominant contribution from Mo-d while the dopant contribution comes from a$^{\prime}_{1}$ and e${'}$. 2V$_\text{2W}$ also induces hybridized spin-down states near E$_\text{F}$ with orbital character similar to that of 2V$_\text{2Mo}$. 2V$_\text{2Mo (or 2W)}$ pushes E$_\text{F}$ close to the valence band edge and provides $\emph{p}$-type spin-polarized character, as shown in Fig.~\ref{fig:dos} (a and g). 2Mn$_\text{2Mo}$ and 2Fe$_\text{2Mo}$ introduce hybridized spin-up mid-gap states with dominant contributions from Mo and Se, respectively. The dopant contribution comes mainly from e${'}$ for 2Mn$_\text{2Mo}$ and e${''}$ for 2Fe$_\text{2Mo}$. Interestingly, 2Mn$_\text{2W}$ stabilizes in AFMC and shows symmetric spin-up and spin-down states with dominant Se-character at E$_\text{F}$ and Mn contribution coming mainly from e${''}$ [Fig.~\ref{fig:dos} (h and k)], while PDOS for 2Fe$_\text{2W}$ shows similar characteristics as 2Fe$_\text{2Mo}$ with spin-up mid-gap states at E$_\text{F}$. Therefore, these doped systems exhibit metallic behavior in one spin channel. On the other hand, the opposite spin channel retains a well-defined band gap and shows a semiconducting character. This half-metallic nature indicates the presence of spin-polarized itinerant carriers and makes these systems highly promising for spintronic applications.

 Understanding the exchange mechanisms responsible for the observed magnetism is crucial for tuning the magnetic properties through dopant selection and designing spintronic materials. Notably, each TM dopant induces a local magnetic moment that couples antiferromagnetically with the surrounding Se-$p$ and Mo (or W)-$d$ states, thereby enabling spin-polarized itinerant carriers to mediate long-range FMC between the TM dopants. This behavior can be observed in the spin-density analysis shown in Fig.~\ref{fig:spin-density}, where spin densities of opposite polarity are localized on the TM dopants and the surrounding atoms. Moreover, structural analysis reveals a TM-Se-TM bond angle of $\approx$ 85$^\circ$, which supports FM superexchange in agreement with the Goodenough-Kanamori-Anderson (GKA) rules~\cite{goodenough1955theory}.
This contribution is expected to be strongest at 1NN separation, where the TM-Se-TM exchange pathway is shortest. Overall, the magnetic exchange interaction arises from a combination of carrier-mediated exchange and Se-$4p$ mediated superexchange. Interestingly, the orbital character and spatial localization of the states near E$_\text{F}$ differ significantly among these three dopants. For 2V$_\text{2Mo}$, the magnetic moment is not strongly localized on the V atoms but is distributed over neighboring Mo atoms [Fig.~\ref{fig:spin-density} (a)], indicating an itinerant character of magnetism. In contrast, 2V$_\text{2W}$ induces magnetic moments that remain largely localized at the V sites [Fig.~\ref{fig:spin-density} (d)], while the W-$d$ dominated states at E$_\text{F}$ suggest that the magnetic interaction is primarily driven by $d$-$d$ exchange.
\begin{figure*}[!ht]
\centering
\includegraphics[width=2\columnwidth]{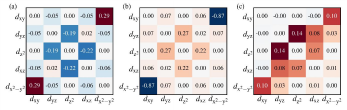}
\caption{Orbital-resolved contributions to MAE (meV per TM dopant) for (a) 2V$_\text{2Mo}$, (b) 2Mn$_\text{2Mo}$, and (c) 2Fe$_\text{2Mo}$ in 2D-MoSe$_{2}$. Each matrix element M$_\text{ij}$ represents MAE due to SOC-induced mixing of d$_\text{i}$ and d$_\text{j}$ orbitals. Positive and negative values correspond to out-of-plane and in-plane MAE, respectively.}
\label{fig:d-res-mae}
\end{figure*} 2Mn$_\text{2Mo}$ and 2Fe$_\text{2Mo}$ exhibit magnetic moments primarily localized at the dopant site, as shown in Fig.~\ref{fig:spin-density} (b and c). For 2Mn$_\text{2Mo}$, Mo-$d$ contribution near E$_\text{F}$ suggests a dominant $d$-$d$ exchange mechanism, while the strong Se-$p$ contribution near E$_\text{F}$ in 2Fe$_\text{2Mo}$ indicates a dominant $p$-$d$ exchange mechanism. 2Mn$_\text{2W}$ and 2Fe$_\text{2W}$ also induce magnetic moments localized at the dopant site [Fig.~\ref{fig:spin-density} (e and f)] with Se-$p$ dominated states at E$_\text{F}$. Overall, the stabilization of FMC, presence of spin-polarized states at E$_\text{F}$, and signatures of carrier-mediated exchange indicate the suitability of these doped systems for spintronics. 

The stability of magnetic ordering against thermal fluctuations is governed by magnetic anisotropy. Controlling the orientation of magnetization and finding the thermally stable axis remain a critical challenge in 2D spintronics and the presence of finite magnetic anisotropy becomes essential. In this regard, we have investigated the preferred magnetization direction using magnetic anisotropy energy (MAE), which quantifies the energy barrier to spin rotation between two degenerate orientations. The MAE is calculated from the total-energy difference between the in-plane and out-of-plane magnetization orientations, including the SOC effect. The MAE is defined as $\mathrm{{MAE} = E^{\parallel} - E^{\perp}} $, where E$^{\parallel}$ and E$^{\perp}$ are the total energies corresponding to the in-plane ($\mathrm{\theta=90^\circ}$) and out-of-plane ($\mathrm{\theta=0^\circ}$) magnetization orientations, respectively. Positive MAE values correspond to perpendicular magnetic anisotropy (PMA), with an out-of-plane easy axis, whereas negative values indicate an in-plane easy axis corresponding to in-plane magnetic anisotropy (IMA). The angular dependence of MAE in the x-z plane is shown in Fig.~\ref{fig:MAE-vs-angle}. Due to the hexagonal symmetry of 2D-MoSe$_\text{2}$ and 2D-WSe$_\text{2}$, x and y directions are equivalent, and the optimized structural parameters of the doped systems do not indicate significant lattice distortion, confirming the preservation of in-plane symmetry. Therefore, separate MAE calculations for the x-z and y-z planes become redundant. We investigate that 2V$_\text{2Mo}$ favors IMA with MAE $\simeq$ $-$3.5 meV, whereas 2Mn$_\text{2Mo}$ and 2Fe$_\text{2Mo}$ stabilize PMA with MAE values $\simeq$ $+$1.5 meV and $+$1.7 meV, respectively. Combined with the exchange energy trends shown in Fig.~\ref{fig:fm-afm-vs-distance}, these results indicate that V and Fe substitution can maintain a stable FM order with fixed spin orientation up to 3NN separation. 2Mn$_\text{2W}$ and 2Fe$_\text{2W}$ favor IMA with MAE $\simeq$ $-$3.19 meV and $-$0.81 meV, respectively; however, they do not exhibit a long-range stable magnetic ordering. Notably, 2V$_\text{2W}$ preserves FM order from 1NN to 3NN separations, but its MAE remains negligible for all orientations. Overall, the combination of robust long-range FMC and large MAE
highlights 2D-MoSe$_\text{2}$ as a more promising host material than 2D-WSe$_\text{2}$ for 2D TMD-based spintronic applications. 

Therefore, we focus on 2V$_\text{2Mo}$, 2Mn$_\text{2Mo}$, and 2Fe$_\text{2Mo}$ to understand the microscopic origin of the MAE. 
The MAE primarily originates from SOC-induced mixing between occupied and unoccupied electronic states. As a result, identifying the relevant orbital coupling is crucial for understanding its microscopic origin. Therefore, we have estimated the contributions of TM$-d$ orbital mixing channels to MAE based on the second-order perturbation theory following the formalism of Wang, Wu, and Freeman\cite{wang1993first}. Since the crystal field splitting of the $d$-orbitals is much larger than the SOC energy, the SOC interaction can be treated as a perturbation. Within this formalism, the MAE originated by SOC is expressed as
\begin{align*}
\text{MAE} = \xi^{2} \Bigg[
&\sum_{\mathrm{o_\uparrow,u_\uparrow}}
\frac{
|\langle \mathrm{o_\uparrow | L_z | u_\uparrow} \rangle|^2
-
|\langle \mathrm{o_\uparrow | L_x | u_\uparrow} \rangle|^2
}{
\mathrm{\varepsilon_{u_\uparrow} - \varepsilon_{o_\uparrow}}
}
\\
+&
\sum_{\mathrm{o_\downarrow,u_\downarrow}}
\frac{
|\langle \mathrm{o_\downarrow | L_z | u_\downarrow} \rangle|^2
-
|\langle \mathrm{o_\downarrow | L_x | u_\downarrow} \rangle|^2
}{
\mathrm{\varepsilon_{u_\downarrow} - \varepsilon_{o_\downarrow}}
}
\\
-&
\sum_{\mathrm{o_\uparrow,u_\downarrow}}
\frac{
|\langle \mathrm{o_\uparrow | L_z | u_\downarrow} \rangle|^2
-
|\langle \mathrm{o_\uparrow | L_x | u_\downarrow} \rangle|^2
}{
\mathrm{\varepsilon_{u_\downarrow} - \varepsilon_{o_\uparrow}}
}
\\
-&
\sum_{\mathrm{o_\downarrow,u_\uparrow}}
\frac{
|\langle \mathrm{o_\downarrow | L_z | u_\uparrow} \rangle|^2
-
|\langle \mathrm{o_\downarrow | L_x | u_\uparrow} \rangle|^2
}{
\mathrm{\varepsilon_{u_\uparrow} - \varepsilon_{o_\downarrow}}
}
\Bigg]
\end{align*}
where o and u denote the occupied and unoccupied states, respectively, with spins $\uparrow$ and $\downarrow$. $\mathrm{L_x}$, and $\mathrm{L_z}$ are components of orbital angular momentum operator along the in-plane (x) and out-of-plane (z) directions, respectively. The non zero angular momentum matrix elements are, ${\langle d_\text{xz} | \mathrm{L_z} | d_\mathrm{yz} \rangle = 1}$, ${\langle d_\mathrm{x^2-y^2}|\mathrm{L_z}|d_\mathrm{xy}\rangle=2}$, ${\langle d_\mathrm{xy}|\mathrm{L_x}|d_\mathrm{xz},d_\mathrm{yz}\rangle=1}$, ${\langle d_\mathrm{z^2}|\mathrm{L_x}|d_\mathrm{xz},d_\mathrm{yz}\rangle=\sqrt{3}}$, and ${\langle d_\mathrm{x^2-y^2}|\mathrm{L_x}|d_\mathrm{xz},d_\mathrm{yz}\rangle=1}$~\cite{wang1993first}. The first two terms correspond to spin-conserving excitations between occupied and unoccupied states of the same spin channel. The last two terms describe spin-flip transitions between occupied and unoccupied states of opposite spin channels. $\xi$ denotes the SOC constant. Figures~\ref{fig:d-res-mae} (a$-$c) represent $5\times5$ SOC coupling matrices of 2V$_\text{2Mo}$, 2Mn$_\text{2Mo}$, and 2Fe$_\text{2Mo}$, respectively. Each non-zero matrix element corresponds to a possible SOC-induced transition between occupied and unoccupied TM-3$d$ orbitals. 
\begin{figure*}[!ht]
\centering
\includegraphics[width=2\columnwidth]{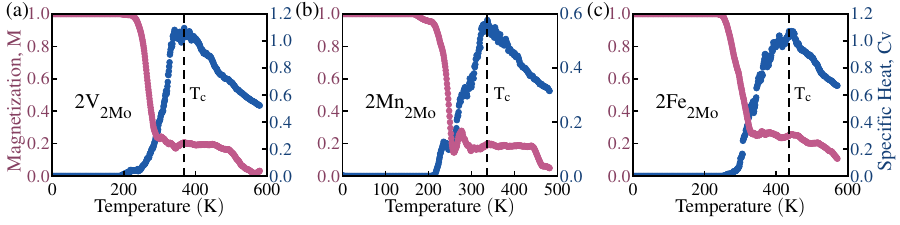}
\caption{The magnetization per spin (M) and the specific heat per spin (C$_\text{v}$) as functions of temperature for (a) 2V$_\text{2Mo}$, (b) 2Mn$_\text{2Mo}$, and (c) 2Fe$_\text{2Mo}$ in 2D-MoSe$_\text{2}$. The Curie temperatures T$\mathrm{_c}$ are estimated from the peaks of the C$_\text{v}$ which are marked by the vertical dashed lines.}
\label{fig:curie_combined}
\end{figure*} 

 For 2V$_\text{2Mo}$ [\ref{fig:d-res-mae} (a)], the mixing between in-plane orbitals (i.e., $d_\mathrm{xy}$ and $d_\mathrm{x^2-y^2}$) mediated by $\mathrm{L_z}$ leads the contribution in PMA. In contrast, the out-of-plane orbitals (i.e., $d_\mathrm{xz}$, and $d_\mathrm{yz}$ mix with $d_\mathrm{z^2}$) mediated by $\mathrm{L_x}$ favor IMA. The competition between these channels results in a net in-plane MAE. In 2Mn$_\text{2Mo}$ [\ref{fig:d-res-mae} (b)], the coupling between $d_\mathrm{xy}$ and $d_\mathrm{x^2-y^2}$ favors IMA, while PMA contributions arise from the coupling of $d_\mathrm{xz}$ and $d_\mathrm{yz}$ with $d_\mathrm{z^2}$. These competing channels nearly compensate each other, resulting in a very small net Mn contribution to MAE. For 2Fe$_\text{2Mo}$ [\ref{fig:d-res-mae} (c)], both $\mathrm{L_x}$ and $\mathrm{L_z}$ driven channels strongly favor PMA. Both in-plane and out-of-plane orbitals mix to favor PMA, with no significant cancellation, resulting in a large PMA originating from Fe atoms. 

Next, we examine the Curie temperature ($\mathrm{T_c}$), which is the key parameter to determine the thermal threshold of spin ordering in magnetic systems. $\mathrm{T_c}$ is calculated using Monte Carlo simulations based on the Ising model with a $50\times50\times1$ supercell. At each temperature, $10^5$ metropolis sweeps have been performed to attain thermal equilibrium, followed by $5\times10^6$ sweeps to obtain thermodynamic properties. The magnetic interactions in 2D systems are generally described by the anisotropic Heisenberg Hamiltonian,
\begin{equation*}
\mathrm{
H = -J \sum_{\langle i,j\rangle} \mathbf{S}_i \cdot \mathbf{S}_j
-\lambda \sum_{\langle i,j\rangle} S_i^{\alpha} S_j^{\alpha}
- D \sum_i \left(S_i^{\alpha}\right)^2
}
\end{equation*}
where, $\mathrm{S_{i}}$ and $\mathrm{{S_j}}$ are spin vectors at sites $\mathrm{i}$ and $\mathrm{j}$ respectively, and $\mathrm{\sum_{\langle i,j\rangle}}$ denotes the summation over nearest neighbor spins. $\alpha$ corresponds to the easy axis of magnetization; J, $\lambda$, and D represent the strength of isotropic Heisenberg exchange interaction, the strength of anisotropic exchange interaction, and single-ion anisotropy, respectively. The large MAE obtained for these doped systems [Fig.~\ref{fig:MAE-vs-angle}] indicates a stable easy axis of magnetization, effectively restricting the spins to align parallel or antiparallel to the easy axis. Therefore, the magnetic behavior can be described by the Ising Hamiltonian \cite{serena1993monte}, $\mathrm{ H = -J \sum_{\langle i,j\rangle} S_i^\alpha S_j^\alpha} $. The resulting Ising Hamiltonian has been used to determine equilibrium spin configurations via the Metropolis algorithm. These configurations are then used to determine thermodynamic quantities: magnetization per spin  M(T), specific heat per spin $\mathrm{C_v(T)}$, staggered magnetization per spin M$_\text{s}$(T) and staggered susceptibility per spin ${\chi_\text{s}}$(T). 
The temperature variation of M(T) and $\mathrm{C_v(T)}$ is shown in Fig.~\ref{fig:curie_combined} for doped 2D-MoSe$_\text{2}$. 
The estimated values of T$_\mathrm{c}$ for 2V$_\text{2Mo}$, 2Mn$_\text{2Mo}$, and 2Fe$_\text{2Mo}$ are $\simeq$ 370 K, 340 K, and 440 K respectively. These T$_\mathrm{c}$ values exceed room temperature (RT), indicating the potential of these DMSs for RT spintronic applications. In particular, 2Fe$_\text{2Mo}$ exhibits the highest T$_\text{c}$, suggesting that the large exchange coupling and magnetic anisotropy stabilizes the FMC against thermal fluctuations. On the other hand, the doped systems in 2D-WSe$_\text{2}$ exhibit considerably lower magnetic transition temperatures. Although the estimated T$_\text{c}$ of 2V$_\text{2W}$ is comparable to RT, the low MAE limits its suitability for RT spintronic applications. For 2Fe$_\text{2W}$, the weak exchange coupling leads to T$_\text{c}$ well below RT.

A notable outcome of our study is the markedly different magnetic behavior of TM doped 2D-MoSe$_2$ and 2D-WSe$_2$ systems. The stronger SOC in WSe$_2$ does not translate into enhanced magnetic anisotropy for TM dopants. Only 2Mn$_\text{2W}$ shows a large in-plane MAE, while 2Fe$_\text{2W}$ shows a substantially smaller MAE, and 2V$_\text{2W}$ exhibits a nearly negligible MAE. More importantly, the exchange coupling parameters are significantly reduced in WSe$_2$ systems, particularly for 2Mn$_\text{2W}$ and 2Fe$_\text{2W}$. As a result, the transition temperatures of the WSe$_2$ systems remain considerably lower than the MoSe$_2$ counterparts, except 2V$_\text{2W}$, which exhibits a comparable exchange coupling parameter.
The predicted half-metallic nature of the doped systems indicates spin-polarized states at the Fermi-level, which are desirable for spin-injection and spin-filtering functionalities. In particular, PMA in 2Fe$_\text{2Mo}$ and 2Mn$_\text{2Mo}$ can be promising for nonvolatile memory elements, whereas IMA in 2V$_\text{2Mo}$ and 2Mn$_\text{2W}$ suggests potential suitability for spin-transport-based device applications. Moreover, the carrier-mediated nature of ferromagnetism opens the scope for tuning the magnetic state via carrier doping. 
\section{CONCLUSION}
\label{conclusion}
 In summary, using DFT$+$U, we have investigated the thermodynamic, electronic, and magnetic properties of $3d$ TM-doped 2D-MoSe$_\text{2}$ and 2D-WSe$_\text{2}$. We show that 3$d$ TM incorporation transforms these non-magnetic semiconductors into half-metallic magnetic systems with spin-polarized states at the Fermi level. Thermodynamic analysis suggest that the clustered dopant configurations are energetically favored over the isolated dopants, particularly under Se-rich growth condition. These clustered dopant configurations in 2D-MoSe$_\text{2}$ simultaneously exhibit long-range FMC, sizable MAE, and high T$_\text{c}$, while none of the dopant configurations in 2D-WSe$_\text{2}$ satisfies these criteria. In particular, 2V$_\text{2Mo}$ and 2Fe$_\text{2Mo}$ emerge as the most promising candidates with estimated T$_\text{c}$ of $\simeq$ 370 K and $\simeq$ 440 K, respectively. Our results reveal that SOC-induced hybridization of dopant $d$ orbitals governs the magnetic anisotropy, giving rise to in-plane magnetic anisotropy of $-$3.5 meV in 2V$_\text{2Mo}$ and perpendicular magnetic anisotropy of 1.7 meV in 2Fe$_\text{2Mo}$. These findings elucidate the mechanisms that govern magnetism in 2D TMD-based DMSs, provide the guidelines for engineering tunable spintronic devices, and identify 3$d$ TM-doped 2D-MoSe$_\text{2}$ as a promising material platform for room temperature spintronic applications. 

\section{Acknowledgements} 
\label{sec:acknowledgements}
 The authors thank Department of Physics, Indian Institute of Technology, Patna for the computational facilities. Akash Singh acknowledges Inspire Faculty Fellowship (IFA 23 MS 194, IVR: 202300014742).

\end{document}